\documentclass{fundam}

\usepackage{url} 
\usepackage{graphicx}

\newcommand{\Int}[1]{$#1$\textnormal{-interval-PCG}}

\usepackage{amsmath,amsfonts,amssymb}
\usepackage{mathrsfs}
\usepackage{soul}

\begin{document}


\setcounter{page}{1}
\publyear{24}
\papernumber{2102}
\volume{193}
\issue{1}

   \versionForARXIV


\title{All Graphs with at Most 8 Nodes are \Int{2}s}

\author{T. Calamoneri\thanks{Address for correspondence:  {\em Sapienza} University of Rome -
                    Computer Science Department, Italy. \newline \newline
                    \vspace*{-6mm}{\scriptsize{Received September 2023; \ accepted October 2024.}}}, \, A. Monti,  \, F. Petroni
 \\
{\em Sapienza} University of Rome - \\
Computer Science Department, Italy\\
calamo@di.uniroma1.it,  monti@di.uniroma1.it}

\maketitle

\runninghead{T. Calamoneri  et al.}{All Graphs with at Most 8 Nodes are \Int{2}s}

\begin{abstract}
A graph $G$ is a {\em multi-interval PCG} if there exist an edge weighted tree $T$ with non-negative real values and disjoint intervals of the non-negative real half-line such that
each node of $G$ is uniquely associated to a leaf of $T$ and
there is an edge between two nodes in $G$ if and only if the weighted distance between their corresponding leaves in $T$ lies within any such intervals.
If the number of intervals is $k$, then we call the graph a \Int{k}; in symbols, $G=$ \Int{k} $(T, I_1, \ldots , I_k)$.

It is known that \Int{2}s do not contain all graphs, and the smallest known graph outside this class has 135 nodes.
Here, we prove that all graphs with at most 8 nodes are \Int{2}s, so doing one step towards the determination of the smallest value of $n$ such that there exists an $n$ node graph that is not a \Int{2}.
\end{abstract}


\section{Introduction}
\label{sec:intro}

A graph $G = (V , E)$ is a {\em pairwise compatibility graph (PCG)} if there exist an edge-weighted tree $T$ with non-negative real values and an interval $I$ of the non-negative real half-line 
such that each node $u \in V$ is uniquely associated to a leaf of $T$ and there is an edge $(u,v) \in E$ if and only if
$d_{T}(u,v) \in I$,
where $d_{T}(u,v)$ is the weighted distance between the leaves associated to $u$ and $v$ in $T$.
In such a case, we say that $G$ is a PCG of $T$ for $I$;
in symbols, $G = PCG(T , I)$.

\eject
The concept of pairwise compatibility was introduced in \cite{Kearney2003} in a computational biology context and it was conjectured that every graph is a PCG.
The conjecture has then been confuted in \cite{Yanhaona2010}, where the first examples of graphs that are not PCGs were provided.
Since then, much work has been done in order to understand which graphs are PCGs ({\em e.g.} see the surveys \cite{Calamoneri2016,Rahman2020}).

Knowing that not all graphs are PCGs, a natural generalization of PCGs has been introduced in \cite{Ahmed2017}:
a graph $G$ is a {\em multi-interval PCG} if there exist an edge weighted tree $T$ with non-negative real values and disjoint intervals of the non-negative real half-line
such that each node of $G$ is uniquely associated to a leaf of $T$
and there is an edge between two nodes in $G$ if and only if the weighted distance between their corresponding leaves in $T$ lies within any such intervals.
If the number of intervals is $k$, then we call the graph a \Int{k}; in symbols, $G=$ \Int{k} $(T, I_1, \ldots , I_k)$.

When $k= 1$, the \Int{1} class coincides with the PCG class.
Instead, already when $k= 2$ the graph class of PCGs is strictly contained in the graph class of \Int{2}s; indeed, there exist graphs that are not PCGs but are \Int{2}s: the smallest such graph has 8 nodes \cite{Durocher2015}; other graphs that are not in PCG but are in \Int{2} are
wheels ({\em i.e.}, one universal node connected to all the nodes of a cycle) with at least 9 nodes and a restricted subclass of series-parallel graphs \cite{Ahmed2017}.

\begin{figure}[!b]
\vspace*{-.8cm}
\centering
\includegraphics[scale=0.49]{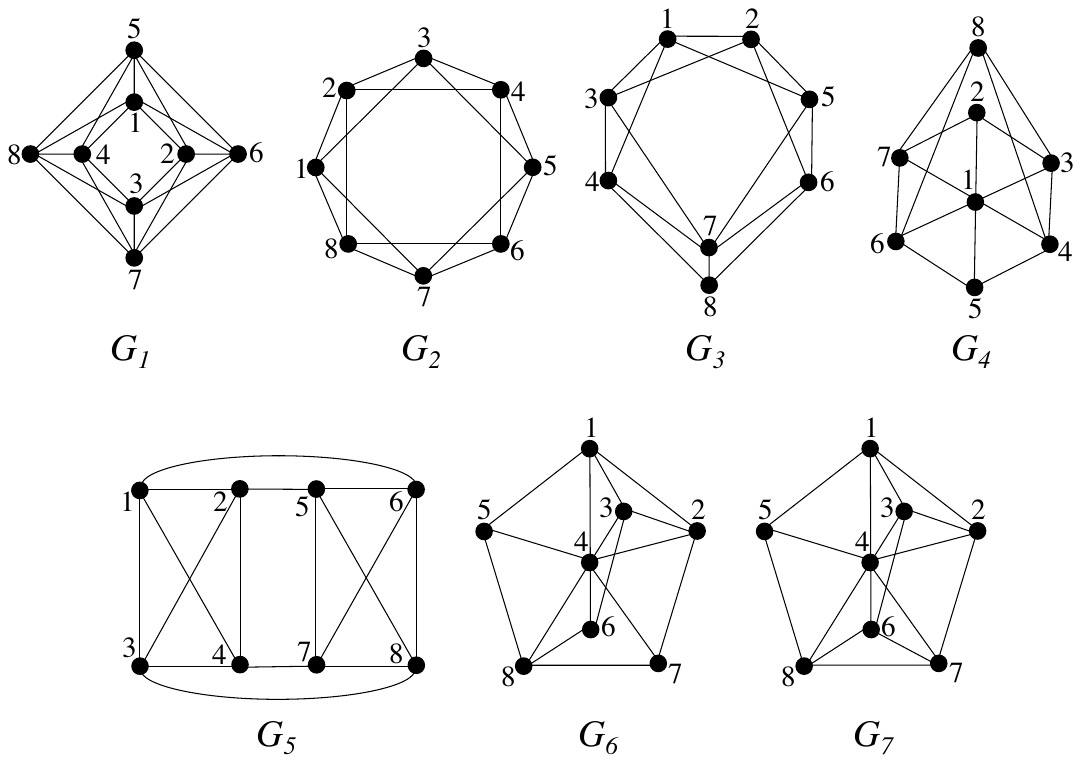}
\vspace*{-8.1cm}
\label{fig.nonPCG}
\caption{The seven graphs with 8 nodes that are not PCGs. \vspace*{-1mm}
}
\end{figure}

It is also known that \Int{2}s do not contain all graphs and the smallest known graph outside this class has 135 nodes \cite{nostro};
moreover, it is known that some classes of graphs, such as grid graphs, are in \Int{2}, whatever their size \cite{PPR2023}.
A natural question is which is the smallest value of $n$ such that there exists an $n$-node graph that is not a \Int{2}.
In this paper, we take one step in this direction:
it is known that all graphs with at most 7 nodes are PCGs \cite{Calamoneri2012} and hence \Int{2}s;
here we show that all graphs with 8 nodes are \Int{2}s, and to do it, it is enough to focus on the 8-node graphs that are not PCGs.
Note that we are working analogously as for PCGs.
Indeed,
the first graph recognized to be outside the PCG class had 15 nodes \cite{Yanhaona2010};
in order to find the smallest graph not in PCG,
Phillips \cite{Phillips} first proved in an exhaustive way that all graphs with less than five nodes are PCGs, then the result was extended to all graphs with at most seven nodes \cite{Calamoneri2012} and later to all bipartite graphs on eight nodes \cite{Mehnaz2013}.
Finally, Durocher, Mondal, and Rahman \cite{Durocher2015} proved that there exists a graph with 8 nodes that is not a PCG.

There exist only seven graphs $G_1, \ldots G_7$ with 8 nodes that are not PCGs \cite{Azam2021} and are depicted in Figure \ref{fig.nonPCG}. Only $G_1$ is already known to be in \Int{2} \cite{nostro}; here we prove that all the others are \Int{2}s as well.
In fact, our result is summarized in Figure 2, and the rest of the paper is needed to justify this figure.
More in detail, this paper is organized as follows:
in Section 2 we
prove some general results giving sufficient conditions for a graph to be in \Int{(k+1)} if certain subgraphs are in \Int{k}, and we exploit them to prove that $G_4$, $G_6$ and $G_7$ are \Int{2}s;
in Section 3 we generalize some results proved for PCGs to let them work for \Int{2}s, too, and we exploit an ILP model to determine whether a given graph is a \Int{2} with respect to a given tree structure;
from it, we deduce that the remaining 8-node graphs are \Int{2}s.
Finally, we are able to find witness trees and intervals for all the graphs in Figure \ref{fig.nonPCG}; we show them in Figures 
2 and \ref{fig.caterpillar}, observing that they have all the same structure (either complete binary tree or caterpillar).
Section 4 concludes the paper.

\section{$G_4$, $G_6$ and $G_7$ are \Int{2}s.}

To show that graphs $G_4$, $G_6$ and $G_7$ of Figure \ref{fig.nonPCG} are \Int{2}s, we prove two theorems giving sufficient conditions for a graph to be a \Int{2}.

Let $G$ be a graph, a {\em universal node} for $G$ is  a node that is adjacent to all other nodes of $G$; a node is {\em almost universal} for $G$
if it is adjacent to all the nodes in $G$ but one.

We now prove the following general result.

\begin{theorem}
\label{theo:universal}
Let $G$ be a graph with a universal node $u$.
If the subgraph $G'$ of $G$, obtained from $G$ by removing $u$, is  a \Int{k}, then $G$ is a \Int{(k+1)}.
\end{theorem}

\begin{proof}
Let $G'=PCG(T,I_1,\ldots I_k)$ with $I_k=[a,b]$.
We construct a  new tree $T_1$ adding to the tree $T$ a leaf corresponding to $u$ and define a new interval $I_{k+1}$ such that $G=PCG(T_1, I_1, \ldots, I_k, I_{k+1})$.

Let $p$ be the maximum distance between the leaves of $T$ (note that it is not restrictive to assume that $b\leq p$).
We construct $T_1$ by adding to $T$ the new leaf corresponding to $u$ as a child of an internal node $x$ of $T$ and edge $(x,u)$ in $T_1$ is assigned a weight $p+1$.

First note that leaf $u$ in $T_1$ has a distance at least $p+1$  from all the other leaves of $T_1$ and no larger than $2p+1$.

We set $I_{k+1}=[p+1,2p+1]$ and observe that the interval $I_1,\ldots I_k$ and $I_{k+1}$ are disjoint
since $b<p+1$. It is easy to see that $G$ coincides with \Int{(k+1)}$(T_1,I_1,\ldots, I_{k+1})$, that is $G$ is in \Int{(k+1)}.
\end{proof}

\noindent
Note that node 4 in graphs $G_6$ and $G_7$ in Figure \ref{fig.nonPCG} is a universal node and their subgraphs obtained deleting it has seven vertices and are hence PCGs (see \cite{Calamoneri2012}).
The next result straightforwardly follows from this observation and from Theorem \ref{theo:universal}:

\begin{corollary}
Non PCG graphs $G_6$ and $G_7$ are in \Int{2}.
\end{corollary}

In order to handle graph $G_4$, we prove the following general result.

\begin{theorem}
\label{theo:almost}
Let $G$ be a graph with an almost universal node $u$. If the subgraph $G'$ of $G$, obtained from $G$ by removing $u$, is  an \Int{k}, then $G$ is a \Int{(k+1)}.
\end{theorem}
\begin{proof}
Let $G'=PCG(T,I_1,\ldots I_k)$ with $I_i=[a_i, b_i]$, $1\leq i\leq k$.
Starting from tree $T$ we now create a new tree $T_1$ by adding a weight $2$
 to all the edges of $T$ incident to the leaves.
In this way, the distances between any two leaves in $T$ are augmented by exactly $4$ in $T_1$ and hence
$G'=PCG(T_1, I'_1,\ldots I'_k)$ where $I_i=[a_i+4,b_i+4]$ and the weight of each edge incident to a leaf is strictly larger than 1.

Let $v$ be the only node not connected with
the almost universal node $u$ of $G$, and let $x$ be the parent of leaf $v$ in $T_1$; call  $c$ the weight of edge $(v,x)$ in $T_1$. (Note that $c \geq 2$).
Now construct a new tree $T_2$ obtained from $T_1$ deleting edge $(v,x)$ and adding a  dummy internal node $y$ connected with $v$ and $x$, where edge $(v,y)$ has weight $1$ and edge $(y,x)$ has weight $c-1$. Obviously, again $G'=PCG(T_2,I'_1,\ldots,I'_k)$.

%
Let $p$ be the maximum distance between the leaves of $T_2$ and hence it is not restrictive to assume that for the right extreme of $I_k$ is $b_k+4 \leq p$.
Finally construct $T_3$ adding to $T_2$ new leaf $u$ connected to $y$ with an edge of weight $p$.
Note that $u$ has distance $p+1$ from $v$ in $T_3$.
Moreover, the distance in $T_3$ between $u$ and any other leaf is less than $2p$ and at least  $p+2$.
Indeed, the weight of edge $(u,y)$ is $p$ and the weight of the unique incident edge of any other leaf is at least 2.
So, we set $I'_{k+1}=[p+2,2p]$.

Easily $I'_1, \ldots I'_k$ and $I_{k+1}$ are disjoint and $G=(T_3,I'_1,\ldots,I'_k,I'_{k+1})$ and hence it is a \Int{(k+1)}.
\end{proof}

\noindent
Node 1 in graph $G_4$ of Figure \ref{fig.nonPCG} is an almost universal node, and the subgraph of $G_4$ obtained by deleting this node is a PCG having seven nodes \cite{Calamoneri2012}.
So, by Theorem \ref{theo:almost},
we have the following result.

\begin{corollary}
Non PCG graph $G_4$ is \Int{2}.
\end{corollary}

We conclude this section observing that the two theorems proved above could be of help even for graphs with a bigger number of nodes, considering large graphs that are known to be not in PCG, recognizing in them a $k$ long sequence of either universal or almost universal nodes whose removal gives a PCG, and deducing that the original graph is at most a \Int{(k+1)}.

\section{$G_2$, $G_3$ and $G_5$ are \Int{2}s}

In order to complete the proof that all graphs with at most 8 nodes are \Int{2}s, we now generalize to \Int{2}s some known results proved for PCGs in order to prove that all graphs with at most 7 nodes are PCGs \cite{Calamoneri2013}.
More precisely, Lemma \ref{le:full_binary}
will allow us to check
the \Int{2} property of a graph only on a special tree structure instead of all possible $n$ leaf trees; Lemma \ref{le:int} will allow us to consider only integer edge-weights and interval extremes instead of real values.
The conjunction of these two simplifications makes effective the ILP model we use to solve our problem.

A {\em full binary tree} is a tree whose all internal nodes, except for its root, have a degree exactly 3.

A {\em caterpillar} is a tree for which any leaf is at a distance exactly one from a central path called spine.
Caterpillars as those depicted in Figure \ref{fig.caterpillar} can be considered as special full binary trees (electing any internal node as root), with the exception that the root has degree 3.

In \cite{Calamoneri2013} it is proved that, given any $n$ leaf tree $T$, there exists an edge-weighted full binary tree $T^*$ on the same leaf set with the property that, for any pair of leaves, their distance in $T$ and in $T'$ is the same.

\noindent
From this fact, the following result follows.

\begin{lemma}
\label{le:full_binary}
Let $G$ be a \Int{2}; then there exists an edge-weighted full binary tree $T^*$ and two disjoint intervals $I_1$ and $I_2$ such that $G=$ \Int{2} $(T^*, I_1, I_2)$.
\end{lemma}

The following result, deduced from \cite{CMPS13}, allows us to use only integer values both for the edge-weight function and for the extremes of the intervals.

\begin{lemma}
\label{le:int}
Let $G$ be a \Int{2}; then there exists an edge-weighted tree $T'$ with integer values and two disjoint intervals $I_1$ and $I_2$ with integer extremes such that $G=$ \Int{2} $(T', I_1, I_2)$ and the tree structure does not change.
\end{lemma}

So, given a \Int{2}, Lemma \ref{le:full_binary} guarantees that one of its witness trees is a full binary tree and Lemma \ref{le:int} ensures that its edge-weight function and the two interval use integer values.

\smallskip

In \cite{Calamoneri2013} an ILP model to determine whether a given graph is a PCG with respect to a given tree structure is provided.
More precisely, given a graph $G$ and an unweighted tree $T$, the model determines (if they exist) an edge-weight function $w$ for $T$,
 a bijective mapping $\sigma$ between the node set $V$ of $G$ and the set  of the leaves of $T$, and an interval $I$ of the non-negative real half-line, such that $G=PCG(T', I)$ where $T'$ is the edge-weighted tree obtained from $T$ introducing weight $w$ and mapping $\sigma$.
After this result, in other papers \cite{Azam2021,Azam2021-9}, the ILP model has been made more efficient in order to work on larger graphs.
Since we handle few small graphs, here we adopt the first ILP model and opportunely modify it in order to let it work for recognizing whether a graph belongs to the wider class of \Int{2}s.
We do not describe here our ILP model because the only change we made to the original one consisted of adapting the constraints to add the second interval.
Nevertheless, for the sake of completeness, we list in the Appendix the constraints we exploited.
It is worth mentioning that the same model could be further extended to a larger number of integers $k$ and check whether a graph belongs to \Int{k}s, although the number of variables and constraints grows.

Finally, note that we are able to exploit such an ILP model because
the \Int{2} property of a graph can be checked only on full binary trees instead of all possible $n$ leaf trees and only for integer edge-weights and intervals with integer extremes.

We apply this ILP model to $G_2$, $G_3$, and $G_5$ and get the following result.

\begin{corollary}
Non PCG graphs $G_2$, $G_3$ and $G_5$ are \Int{2}s.
\end{corollary}

Since the solution of the ILP model produces an edge-weighted tree and two intervals witnessing that the input graph is \Int{2}, we apply it to all 7 graphs of Figure \ref{fig.nonPCG} with two special tree structures: complete binary trees and caterpillars.
In both cases, we obtain the witness edge-weighted trees and the pairs of intervals for all the graphs: they are depicted in Figures 
2 and \ref{fig.caterpillar}.
This was not obvious, indeed in \cite{Calamoneri2012} the authors found that all 7 node graphs are PCGs, all of them except one (the wheel) having as witness tree a caterpillar.
%

\begin{figure}[!ht]
\vspace*{-0.5cm}
\hspace*{1cm}
\includegraphics[scale=0.64]{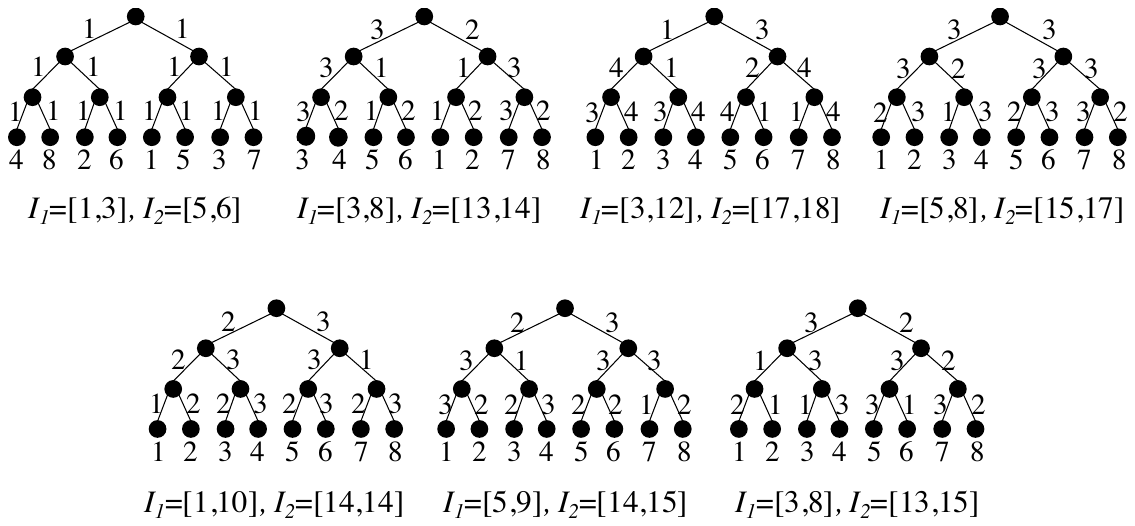}
\vspace*{-12.5cm}
\label{fig.alberi}
\caption{The seven witness complete binary trees with 8 leaves and the corresponding intervals.
}
\end{figure}

\begin{figure}[!ht]
\vspace*{-0.5cm}
\hspace*{1cm}
\includegraphics[scale=0.64]{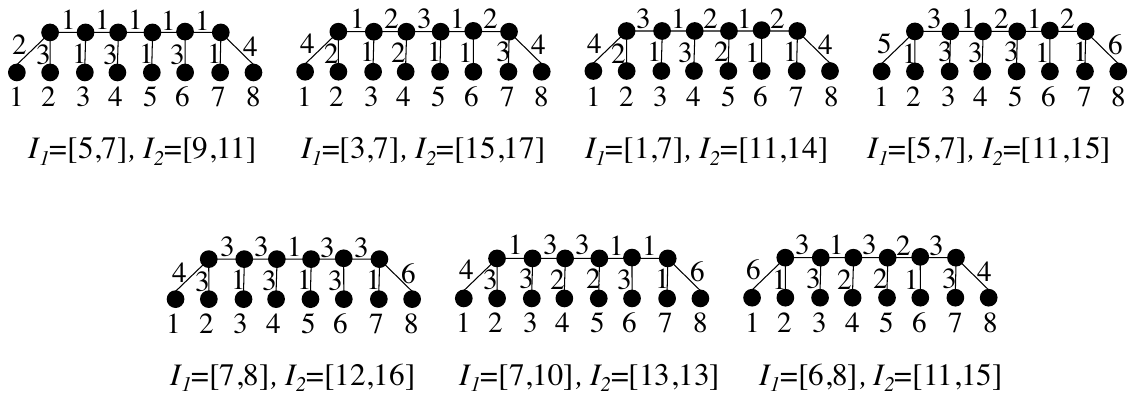}
\vspace*{-14cm}
\label{fig.caterpillar}
\caption{The seven witness caterpillars with 8 leaves and the corresponding intervals.
}
\end{figure}

\section{Conclusions}

It is known that all graphs with at most 7 nodes are PCGs and hence also \Int{2}s.
In this note, we proved that all 8-node graphs are \Int{2}s.

We conclude by observing that 9-node graphs that are not PCGs are much more than seven.
In fact, in \cite{Azam2021-9} only {\em minimal PCGs}  ({\em i.e.}, each of their induced subgraphs is a PCG) with 9 nodes are enumerated, and they are 1494.
So, in order to prove (or disprove) that all 9-node graphs are \Int{2}s, it does not seem possible to apply our approach, even resorting to a most efficient ILP model, as the ones described in \cite{Azam2021,Azam2021-9}.
Instead, it would be interesting to check whether the ILP model extended to larger values of $k > 2$ could be useful to check which graphs are in \Int{k}. Nevertheless, before doing it, we should be able to detect small graphs that are not \Int{2}, and this is currently one of the main open problems on this topic.

\newpage

\section*{Appendix}

Here, we list the constraints we exploited. Explanations are omitted since they can be found in \cite{Calamoneri2013} without any substantial change. We only highlight where our variables and constraints are different with respect to the original ones.

\medskip
Given an $n$ node graph $G = (V, E)$ and an $n$ leaf tree $T=(V', A)$ whose leaf set is denoted by $F \subseteq V'$:
\begin{itemize}
\item $\sigma: V \to F$ is a bijective mapping between the nodes of $G$ and the leaves of $T$;
\item $w(a)$ is the integer weight associated to edge $a \in A$ of $T$;
\item $I_1=[m_1, M_1]$ and $I_2=[m_2, M_2]$ are two intervals with integer extremes, and without loss of generality $m_1 < M_1 < m_2 < M_2$.
\end{itemize}
Denote by $\hat{E} = \{ (i, j) \in V \times V \;:\; i < j \}$ and by $\tilde{F} = \{ (u, v) \in F \times F \;:\; u < v \}$;
since the shape of $T$ is fixed, for each $(u,v) \in \tilde{F}$ let $A(u,v) \subseteq A$ define the unique path between the leaf $u$ and the leaf $v$ in $T$.
Introduce the classical (binary) assignment variables:
\[
 x_{iu} =
 \left\{\begin{array}{ll}
         1 & \mbox{if } \sigma(i)=u\\[0.1cm]
         0 & \mbox{otherwise}
        \end{array}\right.
\]
for all $n^2$ pairs $(i,u) \in V \times F$, together with the assignment constraints:
\begin{equation}
 \textstyle
 \sum_{i \in V} x_{iu} = 1, u \in F \hspace{1cm} \mbox{ and } \hspace{1cm}
 \sum_{u \in F} x_{iu} = 1, i \in V.
 \label{eq:ass}
\end{equation}
For each $(u,v) \in \tilde{F}$ introduce binary variables:
\[
 y^{(1)}_{uv} =
 \left\{\begin{array}{ll}
         1 & \mbox{if } (\sigma^{-1}(u),\sigma^{-1}(v)) \in E \mbox{ thanks to }I_1 \\[0.1cm]
         0 & \mbox{otherwise.}
        \end{array}\right.
 \;\;
\]
\[
 y^{(2)}_{uv} =
 \left\{\begin{array}{ll}
         1 & \mbox{if } (\sigma^{-1}(u),\sigma^{-1}(v)) \in E \mbox{ thanks to }I_2 \\[0.1cm]
         0 & \mbox{otherwise.}
        \end{array}\right.
 \;\;
\]
together with the following constraints:
\begin{equation}
 \textstyle
 y^{(1)}_{uv}+y^{(2)}_{uv} \geq x_{iu} + x_{jv} - 1 \mbox{ if } (i,j) \in E \hspace{1cm} \mbox{ and } \hspace{1cm}
 y^{(1)}_{uv}+y^{(2)}_{uv} \leq 2 - x_{iu} - x_{jv}
 \mbox{ if } (i,j) \notin E.
\end{equation}
The following indicator constraints must hold: \\
\begin{equation}
 \begin{array}{lcl}
  y^{(1)}_{uv} = 1 & \to & \sum_{a \in A(u,v)} w(a) \leq M_1 \\[0.1cm]
  y^{(1)}_{uv} = 1 & \to & \sum_{a \in A(u,v)} w(a) \geq m_1
 \end{array}
\end{equation}
\begin{equation}
 \begin{array}{lcl}
  y^{(2)}_{uv} = 1 & \to & \sum_{a \in A(u,v)} w(a) \leq M_2 \\[0.1cm]
  y^{(2)}_{uv} = 1 & \to & \sum_{a \in A(u,v)} w(a) \geq m_2
 \end{array}
\end{equation}
Note that, in the original ILP model, only one kind of $y_{uv}$ variables were necessary and that constraints (2), (3) and (4) work because, whenever $(\sigma^{-1}(u),\sigma^{-1}(v)) \in E$, $y^{(1)}_{uv}=1 \iff y^{(2)}_{uv}=0$.
\\
We introduce three further binary variables for each $(u,v) \in \tilde{F}$:
\[
\hspace*{-1.7cm}
 y_{uv}^+ =
 \left\{\begin{array}{ll}
         1 & \mbox{if } (\sigma^{-1}(u),\sigma^{-1}(v)) \notin E
             \mbox{ and } \sum_{a \in A(u,v)} w(a) \geq M_2 + 1 \\[0.1cm]
         0 & \mbox{otherwise}
        \end{array}\right.
\]
\[
 y_{uv}^{\pm} =
 \left\{\begin{array}{ll}
         1 & \mbox{if } (\sigma^{-1}(u),\sigma^{-1}(v)) \notin E
             \mbox{ and } M_1 + 1 \leq \sum_{a \in A(u,v)} w(a) \leq m_2 -1 \\[0.1cm]
         0 & \mbox{otherwise}
        \end{array}\right.
\]
\[
\hspace*{-1.7cm}
 y_{uv}^- =
 \left\{\begin{array}{ll}
         1 & \mbox{if } (\sigma^{-1}(u),\sigma^{-1}(v)) \notin E
             \mbox{ and } \sum_{a \in A(u,v)} w(a) \leq m_1 - 1 \\[0.1cm]
         0 & \mbox{otherwise}.
        \end{array}\right.
\]
together with the following constraints:
\begin{equation}
 1 - y^{(1)}_{uv}-y^{(2)}_{uv} = y_{uv}^+ + y_{uv}^{\pm} + y_{uv}^-
 \label{eq:lnkng3}
\end{equation}
Finally, the following further indicator constraints must hold:
\begin{equation}
 \begin{array}{lcl}
  y_{uv}^+ = 1 & \to & \sum_{a \in A(u,v)} w(a) \geq M_2 + 1 \\[0.1cm]
    y_{uv}^{\pm} = 1 & \to & M_2 +1 \leq \sum_{a \in A(u,v)} w(a) \leq m_1 - 1 \\[0.1cm]
  y_{uv}^- = 1 & \to & \sum_{a \in A(u,v)} w(a) \leq m_1 - 1
 \end{array}
 \label{eq:indminus}
\end{equation}
Also in this case, the original ILP model was simpler, and only two kinds of these variables were necessary ($y^+_{uv}$ and $y^-_{uv}$).

\medskip\noindent
We conclude by highlighting that we do not need to specify any objective function because we are only interested in the feasibility of the integer system.

\end{document}